\documentstyle[sprocl,psfig]{article}

\def\beq{\begin{equation}}
\def\eeq{\end{equation}}
\def\bea{\begin{eqnarray}}
\def\eea{\end{eqnarray}}

\begin{document}

\title{\vspace*{-1.0cm} \hfill {\rm MKPH-T-99-22}\\ \vspace{0.5cm}
Pion Electroproduction on the Nucleon and the Generalized GDH
  Sum Rule}

\author{L. Tiator, D. Drechsel and S.S. Kamalov\footnote{permanent address:
Laboratory of Theoretical Physics, JINR Dubna, 141980 Moscow region,
Russia.}}

\address{Institut f\"{u}r Kernphysik, Johannes Gutenberg-Universit\"{a}t,
J.~J.~Becher-Weg 45, D-55099 Mainz, Germany\\
E-mail: tiator@kph.uni-mainz.de}


\maketitle

\abstracts{ We present predictions for the spin structure functions of
the proton and the neutron in the framework of a unitary isobar model
for one-pion photo- and electroproduction. Our results are compared
with recent experimental data from SLAC. The first moments of the
calculated structure functions fulfill the Gerasimov-Drell-Hearn and
Burkhardt-Cottingham sum rules within an error of typically 5-10\% for
the proton. For the neutron target we find much bigger deviations, in
particular the sum rule for $I_1(0)+I_2(0)$ is heavily violated.}

\section{Introduction}\label{sec:intro}
The spin structure of the nucleon in the resonance region is of
particular interest to understand the rapid transition from resonance
dominated coherent processes to incoherent processes of deep inelastic
scattering (DIS) off the constituents. By scattering polarized lepton
beams off polarized targets, it has become possible to determine the
spin structure functions $g_1$ and $g_2$. The results of the first
experiments at CERN~\cite{Ash89} and SLAC~\cite{Bau83} sparked
considerable interest in the community, because the first moment of
$g_1$, $\Gamma_1=\int^1_0g_1(x)dx$, was found to be substantially
smaller than expected from the quark model, in particular from the
Ellis-Jaffe sum rule~\cite{Ell74}.

Here we present the results of the recently developed Unitary Isobar
Model (UIM, Ref.~\cite{Dre98a})\footnote{an online version (MAID) is
available in the internet at www.kph.uni-mainz.de/T/maid/} for the spin
asymmetries, structure functions and relevant sum rules in the
resonance region. This model describes the presently available data for
single-pion photo- and electroproduction up to a total $cm$ energy
$W_{\rm max}= 1.7$ GeV and for $Q^2\le$ 2 (GeV/c)$^2$. It is based on
effective Lagrangians for Born terms (background) and resonance
contributions, and the respective multipoles are constructed in a
gauge-invariant and unitary way for each partial wave.  The eta
production is included in a similar way~\cite{Kno95}, while the
contribution of more-pion and higher channels is modeled by comparison
with the total cross sections and simple phenomenological assumptions.

\section{Formalism}\label{sec:form}

The differential cross section for exclusive electroproduction of
mesons from polarized targets using polarized electrons, e.g. $\vec
p(\vec e,e'\pi^0)p$ can be parametrized in terms of 18 response
functions\cite{Dre92}, a total of 36 is possible if in addition also
the recoil polarization is observed. Due to the azimuthal symmetry most
of them vanish by integration over the angle $\phi$ and only 5 total
cross sections remain. The differential cross section for the electron
is then given by
\beq \frac{d\sigma}{d\Omega\ dE'} = \Gamma\sigma
(\nu,Q^2)\,, \label{eq1}
\eeq
\beq
\sigma =\sigma_T+\epsilon\sigma_L+
P_y\sqrt{2\epsilon(1+\epsilon)}\
         \sigma_{LT}+
         hP_x\sqrt{2\epsilon(1-\epsilon)}\
         \sigma_{LT'}+hP_z\sqrt{1-\epsilon^2}\sigma_{TT'}\, ,
\label{eq2}
\eeq
where $\Gamma$ is the flux of the virtual photon field and the
$\sigma_i,$ $i=L$, $T$, $LT$, $LT'$, $TT'$, are functions of the $lab$
energy of the virtual photon $\nu$ and the squared four-momentum
transferred $Q^2$. These response functions can be separated by varying
the transverse polarization $\epsilon$ of the virtual photon as well as
the polarizations of the electron ($h$) and proton ($P_z$ parallel,
$P_x$ perpendicular to the virtual photon, in the scattering plane and
$P_y$ perpendicular to the scattering plane). In particular,
$\sigma_{T}$ and $\sigma_{TT'}$ can be expressed in terms of the total
cross sections for excitation of hadronic states with spin projections
$3/2$ and $1/2$: $\sigma_{T}=(\sigma_{3/2}+\sigma_{1/2})/2$ and
$\sigma_{TT'}=(\sigma_{3/2}-\sigma_{1/2})/2.$

In inclusive electron scattering  $\vec e+\vec N\rightarrow X$, only 4
cross sections $\sigma_T$, $\sigma_L$, $\sigma_{LT'}$ and
$\sigma_{TT'}$ appear, the fifth cross section, $\sigma_{LT}$, vanishes
due to unitarity when all open channels are summed up. The individual
channels, however, give finite contributions.

The relations between the $\sigma_i$ and the quark structure functions
$g_1$ and $g_2$ can be read off the following equations, which define
possible generalizations of the Gerasimov-Drell-Hearn (GDH)
integral~\cite{Ger65} and the Burkhardt-Cottingham (BC) sum rule~\cite{Bur70},
\begin{eqnarray}
I_{GDH}(Q^2) &=& \frac{2m^2}{Q^2}\int_{0}^{x_0}
     \left (g_1(x,Q^2)-\gamma^2 g_2(x,Q^2)\right )\ dx \nonumber
\\
&=&\frac{m^2}{8\pi^2\alpha}\int_{\nu_0}^{\infty}
           (1-x)
           \left (\sigma_{1/2}-\sigma_{3/2}
           \right )\ \frac{d\nu}{\nu}\
\\
I_1(Q^2) &=& \frac{2m^2}{Q^2}\int_{0}^{x_0}g_1(x,Q^2)\ dx \nonumber
\\
&=&\frac{m^2}{8\pi^2\alpha}\int_{\nu_0}^{\infty}
           \frac{1-x} {1+\gamma^2}
           \left (\sigma_{1/2}-\sigma_{3/2}
           -2\gamma\,\sigma_{LT'}\right )\ \frac{d\nu}{\nu}\
\\
I_2(Q^2) &=& \frac{2m^2}{Q^2}\int_{0}^{x_0}g_2(x,Q^2)\ dx \nonumber
\\
&=&  \frac{m^2}{8\pi^2\alpha}\int_{\nu_0}^{\infty}
           \frac{1-x} {1+\gamma^2}
           \left (\sigma_{3/2}-\sigma_{1/2}
           -\frac{2}{\gamma}\,\sigma_{LT'}\right )\ \frac{d\nu}{\nu}\ ,
\\
I_3(Q^2) &=& \frac{2m^2}{Q^2}\int_{0}^{x_0}(g_1(x,Q^2)+g_2(x,Q^2))\ dx \nonumber
\\
&=&  -\frac{m^2}{4\pi^2\alpha}\int_{\nu_0}^{\infty}
           \frac{1-x}{Q}\,\sigma_{LT'}\ d\nu\ = I_1+I_2 ,
\label{eq3}
\end{eqnarray}
where $\gamma=Q/\nu$ and $x=Q^2/2m\nu$ the Bjorken scaling variable,
with $x_0$ ($\nu_0$) referring to the inelastic threshold of one-pion
production. Since $\sigma_{LT'}={\cal O}(Q)$, the real photon limit of
the integral $I_1$  is given by the GDH sum rule
$I_1(0)=I_{GDH}(0)=-\kappa_N^2/4,$ with $\kappa_N$ the anomalous
magnetic moment of the nucleon. At large $Q^2$ the structure functions
should depend only on $x,$ i.e. $I_1\rightarrow 2m\Gamma_1/Q^2$ with
$\Gamma_1=\int g_1(x){\rm d}x=$const. In the case of the proton, all
experiments for $Q^2> 1$GeV$^2$ yield $\Gamma_1>0.$ Therefore, a strong
variation of $I_1(Q^2)$ with a zero-crossing at $Q^2<1$ GeV$^2$ is
required in order to reconcile the GDH sum rule with the measurements
in the DIS region. The third integral of Eq.~(5) is constrained by the
BC sum rule, which requires that the inelastic contribution for $0 < x
<x_0 $ equals the negative value of the elastic contribution, i.e.
\begin{equation}
I_2(Q^2) = \frac{2m^2}{Q^2}\int_{0}^{x_0}g_2(x,Q^2)\ dx
=\frac{1}{4}\frac{G_M(Q^2)-G_E(Q^2)}{1+Q^2/4m^2}\,G_M(Q^2)\,,
\label{eq7}
\end{equation}
where $G_M$ and $G_E$ are the magnetic and electric Sachs form factors
respectively. At large $Q^2$ the integral vanishes as $Q^{-10}$, while
at the real photon limit $I_2(0)=\kappa_N^2/4+e_N\kappa_N/4$, the two
terms on the right hand side corresponding to the contributions of
$\sigma_{TT'}$ and $\sigma_{LT'}$ respectively. Finally, Eq. (6)
defines an integral $I_3(Q^2)$ as the sum of $I_1(Q^2)$ and $I_2(Q^2)$
and is given by the unweighted integral over the longitudinal
transverse interference cross section $\sigma_{LT'}$. At the real
photon point this integral is given by the GDH and BC sum rules,
$I_3(0)=e_N\kappa_N/4$. In particular this vanishes for the neutron
target.

\section{Unitary Isobar Model}
Our calculation for the response functions $\sigma_i$ is based on the
Unitary Isobar Model (UIM) for one-pion photo- and electroproduction of
Ref.~\cite{Dre98a}. The model is constructed with effective
phenomenological Lagrangians for Born terms, vector meson exchange in
the $t$ channel (background), and the dominant resonances up to the
third resonance region. For each partial wave the multipoles satisfy
gauge invariance and unitarity. As in any realistic model a special
effort is needed to describe the $s$-channel multipoles $S_{11}$ and
$S_{31}$. Even close at threshold these multipoles pick up sizeable
imaginary parts that cannot be explained by nucleon resonances. In fact
the $S_{11}(1535)$, $S_{11}(1650)$ and the $S_{31}(1620)$ play only a
minor role for the complex phase of the $E_{0+}$ multipoles even at
higher energies. The main effect arises from pion rescattering. This we
can take into account by $K$-matrix unitarization. Furthermore we
introduce a mixing of pseudoscalar (PS) and pseudovector (PV) $\pi NN$
coupling in the form
\begin{equation}
{\cal L}_{\pi NN}^{HM}=\frac{\Lambda_m^2}{\Lambda_m^2+q_0^2} {\cal
L}_{\pi NN}^{PV}+\frac{q_0^2}{\Lambda_m^2+q_0^2} {\cal L}_{\pi
NN}^{PS}\,,
\end{equation}
where  $q_0$ is the asymptotic pion momentum in the $\pi N$ $cm$ frame
which depends only on $W$ and is not an operator acting on the pion
field. From the analysis of the $M_{1-}^{(3/2)}$ and $E_{0+}^{(3/2)}$
multipoles we have found that the most appropriate value for the mixing
parameter is $\Lambda_m=450$ MeV. This form satisfies gauge invariance
and chiral symmetry in the low-energy limit and generates in a simple
phenomenological way the effects of pion loops. Expressed in the
invariant electroproduction amplitudes $A_{1-6}(s,t,u)$ we find a
change from the usual amplitudes for PV coupling only in $A_1$,
\beq
A_1^I = A_1^{I,PV} -
\frac{eg}{2m^2}F(q_0^2)(\kappa_V\delta_{I,+}+\kappa_S\delta_{I,0}),
\eeq
where $I$ denotes the isospin component $(+,-,0)$, $\kappa_{V,S}$ are
the isovector and isoscalar anomalous magnetic moments of the nucleon
and $F(q_0^2)=q_0^2/(\Lambda_m^2+q_0^2)$ is a form factor that vanishes
in the chiral limit. Since the modification acts as a contact term and
because of the smallness of $\kappa_S=-0.06$, only the multipoles
$E_{0+}^{(+)}$, $M_{1-}^{(+)}$, $L_{0+}^{(+)}$ and $L_{1-}^{(+)}$ are
really affected.
\begin{figure}[ht]
\label{fig1}
\centerline{\psfig{file=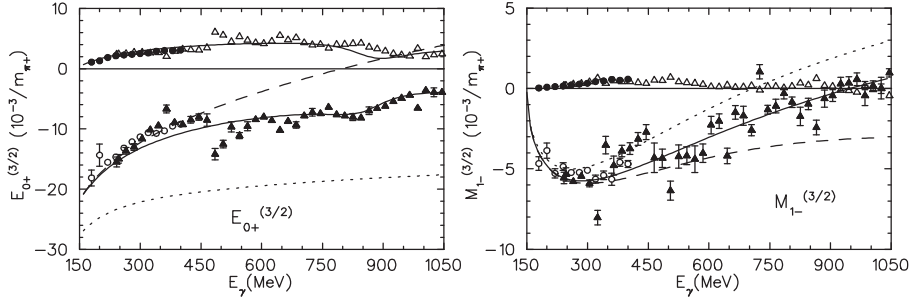,width=12cm,angle=0,silent=}} \caption{
Non-resonant $E_{0+}$ and $M_{1-}$ multipoles calculated with pure
  pseudovector (dashed curves) and pure pseudoscalar (dotted curves)
  $\pi NN$ couplings. The solid curves are the results for the real
  and imaginary parts obtained using the Lagrangian, Eq. (8) and the
  K-matrix unitarization. The open and full circles
  are the real and imaginary parts from the Mainz dispersion
  analysis\protect\cite{HDT}.  The full and open triangles are real
  and imaginary parts from the VPI analysis\protect\cite{VPI97}.
 }
\end{figure}
In Fig. 1 we show a comparison for the non-resonant multipoles
$E_{0+}^{(3/2)}$ and $M_{1-}^{(3/2)}$ with the multipole analysis of
Hanstein et al.\cite{HDT} and the GWU/VPI group\cite{VPI97}. While the
PV Born terms very well describe the multipoles in the threshold
region, they fail to reproduce the experimental multipoles at higher
energies. Furthermore unitarization does not play a very big role in
these non-resonant multipoles, in particular for the $P_{31}$, where
the imaginary part is very small. Therefore, these two multipoles can
ideally be used to fix the free PS-PV mixing parameter $\Lambda_m$. Up
to a total $cm$ energy $W_{\rm max}=1.7$ GeV and for $Q^2\le$ 2
(GeV/c)$^2$ the UIM is able to describe the single-pion
electroproduction channel quite well. However, at higher energies the
contributions from other channels become increasingly important. In the
structure functions $\sigma_{T}$ and $\sigma_{TT'}$ we account for the
$\eta$ and the multi-pion production contributions extracting the
necessary information from the existing data for the total cross
section~\cite{Dre98b}. In Fig. 2 we show the individual channels for
the total cross section at $Q^2=0$ and at $Q^2=0.5 GeV^2$.
\begin{figure}[ht]
\label{fig2}
\centerline{\psfig{file=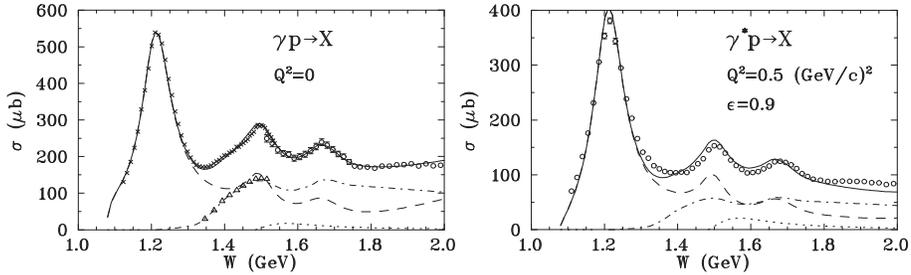,width=12cm,angle=00,silent=}}
\caption{
Total cross sections for photoabsorption and inelastic
  electron scattering on the proton for $\varepsilon=0.9$ and
  $Q^2=0.5\,(GeV/c)^2$.  Dashed, dotted and dash-dotted curves:
  contributions of single-pion, eta and multi-pion channels,
  respectively; solid curves: final result. Experimental data for the
  total cross sections from Refs.\protect\cite{Mac96} (x) and
  \protect\cite{Bra76} ($\circ$), for the two-pion production channels
  from Ref.\protect\cite{Mac96} ($\bigtriangleup$).
 }
\end{figure}

\begin{figure}[ht]
\label{fig3}
\centerline{\psfig{file=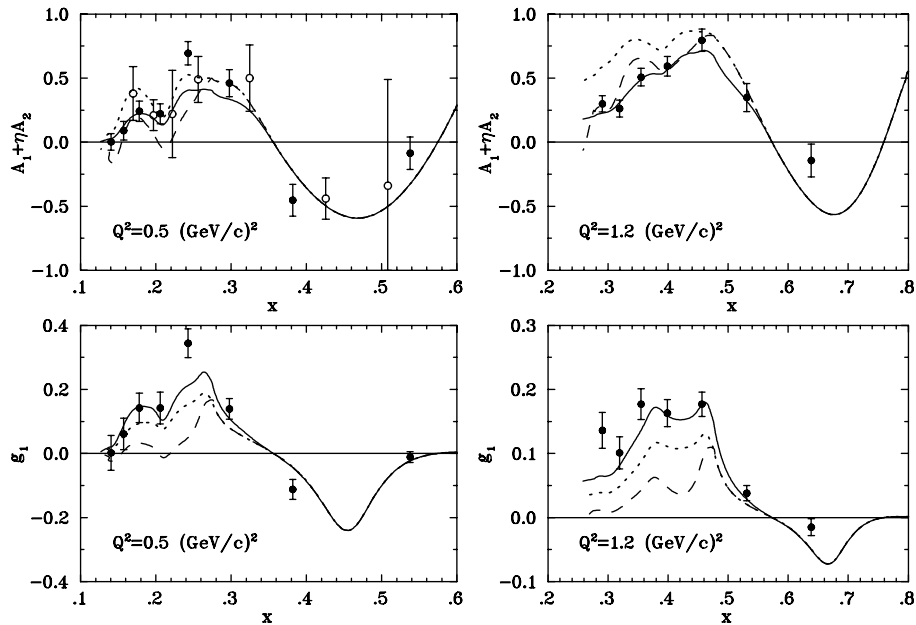,width=12cm,angle=90,silent=}}
\caption{
  The asymmetry $A_1+\eta A_2$ (top) and the spin structure function
  $g_1$ (bottom) as function of $x$ at
  $Q^2=0.5$ and 1.2 $(GeV/c)^2$. Dashed, dotted and solid curves:
  calculations obtained with $1\pi$, $1\pi+\eta$, and
  $1\pi+\eta+n\pi$ contributions, respectively. Data from
  Refs.~\protect\cite{Abe98} ($\bullet$) and \protect\cite{Bau80}
  $(\circ)$.}
\end{figure}
\begin{figure}[ht]
\label{fig4}
\centerline{\psfig{file=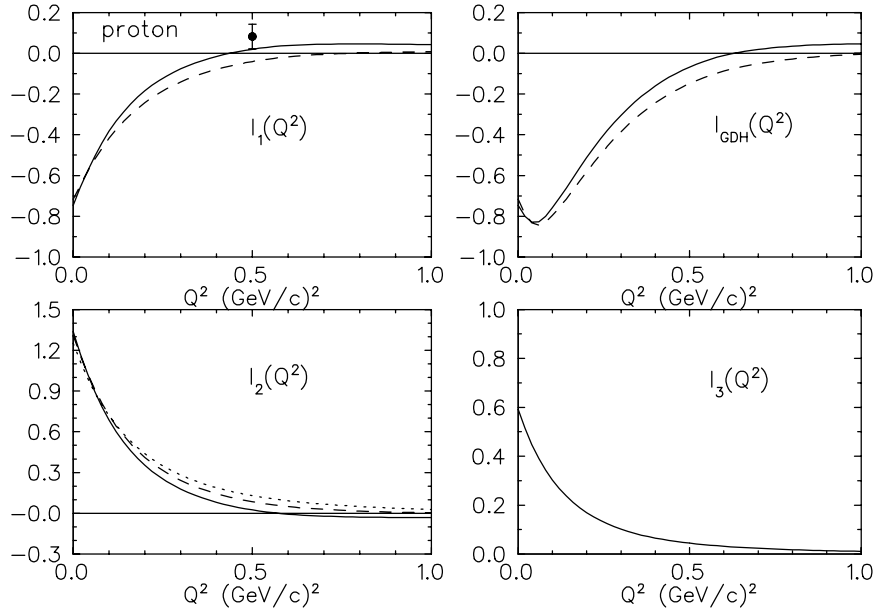,height=8cm,angle=90,silent=}}
\caption{
  The integrals $I_{GDH}$, $I_1$, $I_2$ and $I_3=I_1 + I_2$ as
  functions of $Q^2$, integrated up to
  $W_{{\rm max}}=2$ GeV for the proton target. The dashed
  lines show the contributions from the $1\pi$ channel while the full
  lines include $1\pi + \eta +n\pi$.
  The dotted line is the sum rule prediction of~{Ref.~\protect\cite{Bur70}}.
  The data is from Ref.\protect\cite{Abe98}. }
\end{figure}

\begin{figure}[ht]
\label{fig5}
\centerline{\psfig{file=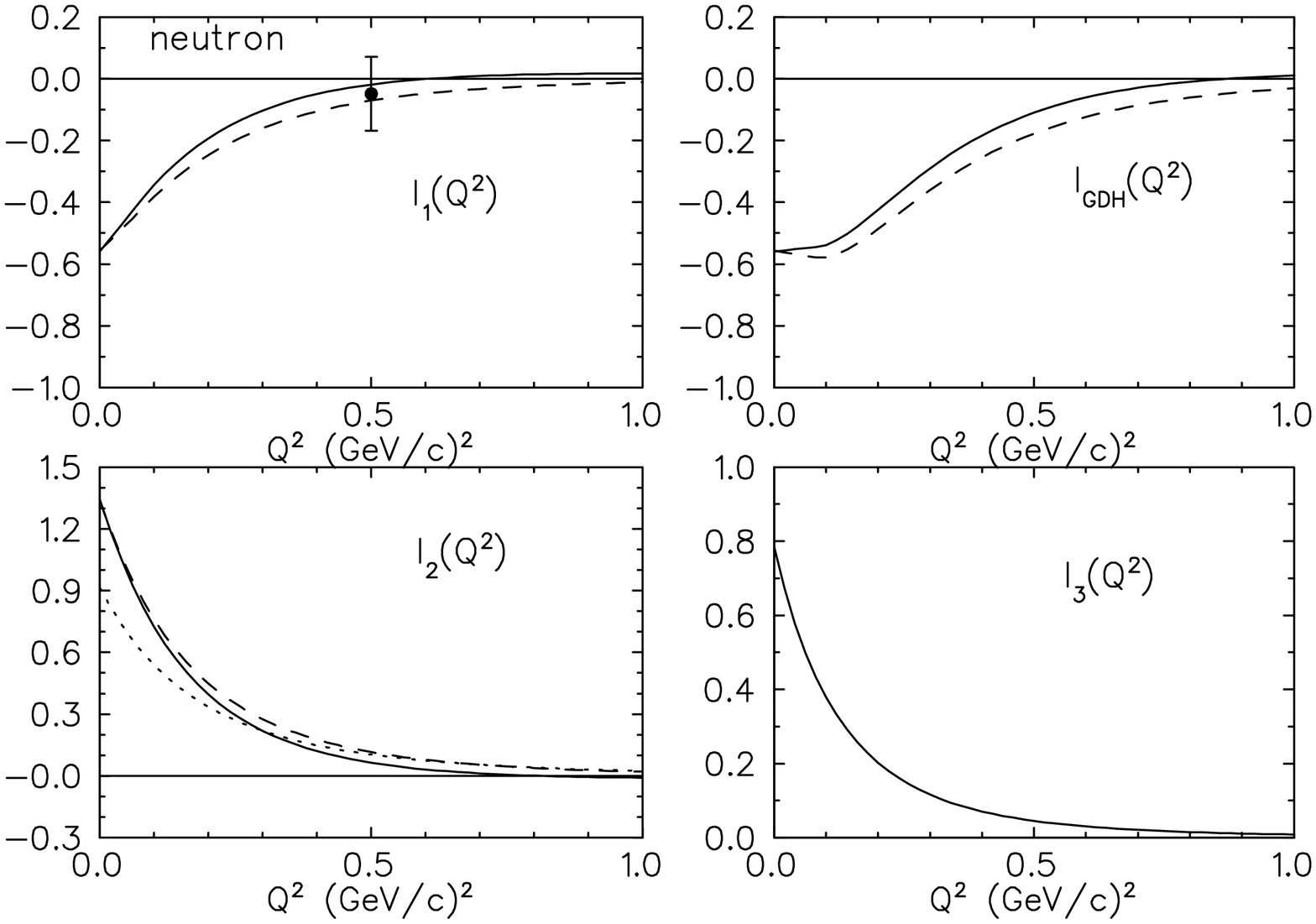,height=8cm,angle=90,silent=}}
\caption{
 The same as Fig. 4 for the neutron target.
  }
\end{figure}
In the upper part of Fig. 3, our results for the asymmetry $A_1+\eta
A_2$ are compared with the data from SLAC~\cite{Abe98}. The asymmetry
is calculated in terms of the virtual photon cross sections by use of
the relations $A_1=-\sigma_{TT'}/\sigma_{T}$ and
$A_2=-\sigma_{LT'}/\sigma_{T}$. We find a reasonable agreement with the
data up to $W=2 $ GeV.  We also note that the contribution of the
$\eta$ channel (dotted curves) leads to a substantial increase of the
asymmetry over a wide energy region. In the lower part of Fig. 3 we
show our results for the structure function $g_1.$  Up to a value of
$W^2=2$ GeV$^2$ (corresponding to $x=0.31$ and $x=0.52$ at $Q^2$=0.5
and 1.2 (GeV/c)$^2$, respectively), the main contribution to $g_1$ is
due to single-pion production. We clearly note the negative structure
above threshold related to excitation of the $\Delta(1232)$ resonance,
while in the second and third resonance regions the contributions from
$\eta$ and multi-pion channels become increasingly important.

\section{Integrals}

In Fig. 4 and Fig. 5 we give our predictions for the integrals
$I_{GDH}(Q^2)$, $I_1(Q^2)$, $I_2(Q^2)$ and $I_3(Q^2)$ in the resonance
region, i.e. integrated up to $W_{{\rm max}}=2$ GeV for the proton and
neutron targets. In the case of the integral $I_1,$ our model is able
to generate the expected drastic change in the helicity structure at
low $Q^2$. We find a zero-crossing at $Q^2=0.75$ (GeV/c)$^2$ if we
include only the one-pion contribution. This value is lowered to 0.52
(GeV/c)$^2$ and 0.45 (GeV/c)$^2$ when we include the $\eta$ and the
multi-pion contributions respectively. The SLAC analysis yield
$I_1=0.1\pm0.06$ at $Q^2= 0.5$ (GeV/c)$^2$, while our result at this
point is slightly positive. This deviation could be ascribed mainly to
two reasons. First, due to a lack of data points in the $\Delta$
region, the SLAC data are likely to underestimate the $\Delta$
contribution. Second, the strong dependence of the zero-crossing on the
multi-pion channels gives rise to uncertainties in our model. A few
more data points in the $\Delta$ region would help to clarify this
situation. Comparing the two generalizations of the GDH sum rule,
$I_1(Q^2)$ and $I_{GDH}(Q^2)$ it can be seen that the slope at $Q^2=0$
depends on the inclusion of the longitudinal contributions. For the
proton target the slope gets even an opposite sign with a pronounced
minimum for $I_{GDH}$ at $Q^2\approx 0.05 (GeV/c)^2$. This behaviour
was also recently obtained in an effective Lagrangian approach
\cite{Schol99}.

Concerning the integral $I_2$, our full result is in good agreement
with the prediction of the BC sum rule.  The deviation is within
$10~\%$ and should be attributed to contributions beyond $W_{\rm
max}=2$ GeV and the uncertainties in our calculation for
$\sigma_{LT'}$. As seen in Eq. (6) the integral $I_3$ depends only on
this $\sigma_{LT'}$ contribution. From the sum rule result a value of
$e_N\kappa_N/4$ is expected at $Q^2=0$, i.e. 0.45 for the proton and
zero for the neutron target. While our value arising entirely from the
$1\pi$ channel gets relatively close to the sum rule result for the
proton, in the neutron case this sum rule is heavily violated. So far
it is not clear where such a large negative contribution should arise
for the neutron target. Either it is due to the high energy tail that
may converge rather slowly for the unweighted integral $I_3$, or the
multi-pion channels could contribute in such a way, while the eta
channel is very unlikely. On the other hand the convergence of the BC
sum rule cannot be given for granted. In fact Ioffe et al. \cite{Iof84}
have argued that the BC sum rule is valid only in the scaling region,
while it is violated by higher twist terms at low $Q^2$. In any case a
careful study of the multi-pion contribution for both proton and
neutron targets will be very helpful, in particular one can expect
longitudinal contributions from the non-resonant background.

In Table 1 we list the contributions of the different ingredients of
our model to the integrals $I_1$, $I_2$ and $I_3$ at the real photon
point, $Q^2=0$, for both protons and neutrons. Our values are
calculated with the upper limit of integration $W_{\rm max}=2$ GeV. At
the photon point the contribution of the $\eta$ and the multi-pion
channels tend to cancel each others. This is no longer the case for
$Q^2 \geq 0.4 (GeV/c)^2$. A complementary analysis to estimate the
non-resonance contribution to the generalized GDH integral was recently
reported in Ref.~\cite{Bianchi}.
\begin{table}[htbp]
\caption{Contributions of the different channels to the integrals
$I_1$, $I_2$ and $I_3=I_1 + I_2$  at the photon point, $Q^2=0$. (upper
part for the proton, lower part for the neutron target) } \vskip 0.3 cm
\begin{tabular}{lrrrrrr}
\hline
 $I_{1,2,3}$ & Born+$\Delta$  & $P_{11},D_{13},...$
& $\eta$ & multi-pion & total & sum rule \\ \hline
 $I_1$ & -0.565 & -0.152 &  0.059 & -0.088 & -0.746 & -0.804\\
 $I_2$ &  1.246 &  0.063 & -0.059 &  0.088 &  1.338 &  1.252\\
 $I_3$ &  0.681 & -0.089 &      0 &      0 &  0.592 &  0.448\\
\hline
 $I_1$ & -0.617 &  0.061 &  0.039 & -0.064 & -0.581 & -0.912\\
 $I_2$ &  1.414 & -0.075 & -0.039 &  0.064 &  1.364 &  0.912\\
 $I_3$ &  0.797 & -0.014 &  0     &      0 &  0.783 &      0\\
\hline
\end{tabular}
\end{table}

\section{Summary}

In summary, we have applied our recently developed unitary isobar model
for pion electroproduction to calculate generalized GDH integrals and
the BC sum rule for both proton and neutron targets. Our results
indicate that both the experimental analysis and the theoretical models
have to be quite accurate in order to fully describe the helicity
structure of the cross section in the resonance region.

While our results agree quite well for the GDH and BC sum rules for the
proton, we find substantial deviations for the neutron target, in
particular the sum rule $I_3(0)\equiv I_1(0)+I_2(0)=0$ is heavily
violated by the contribution from the single-pion channel which is even
larger than in the case of the proton.

Concerning the theoretical description, the treatment of the multi-pion
channels has to be improved with more refined models. On the
experimental side, the upcoming results from measurements with
real~\cite{Ahr92} and virtual photons~\cite{Bur91} hold the promise to
provide new precision data in the resonance region.

\section*{Acknowledgments}
 We would like to thank G.\ Krein, O.\ Hanstein, and B.\ Pasquini for a
fruitful collaboration and J.\ Arends and P.\ Pedroni for useful
discussions on experimental subjects and data analysis. This work was
supported by the Deutsche Forschungsgemeinschaft (SFB 443).

\end{document}